\documentclass[10pt]{wlpeerj}

\usepackage{hyperref}
\usepackage{soul}
\usepackage[normalem]{ulem}

\title{LANTERN: TCR-Peptide Binding Prediction via Large Language Model Representations}

\author[1]{Cong Qi}
\author[1]{Hanzhang Fang}
\author[1]{Siqi Jiang}
\author[1]{Tianxing Hu}
\author[1]{Zhi Wei}
\corrauthor[1]{Zhi Wei}{zhi.wei@njit.edu}

\affil[1]{New Jersey Institute of Technology}

\begin{abstract}

Predicting T-cell receptor (TCR) and peptide-major histocompatibility complex (pMHC) interactions is critical for advancing targeted immunotherapies and personalized medicine. However, existing models often struggle with limited labeled data and poor generalization to novel epitopes. We present \textbf{LANTERN} (Large lAnguage model-powered TCR-Enhanced Recognition Network), a novel deep learning framework that combines pretrained protein and molecular language models with a cross-modality fusion mechanism. Specifically, LANTERN encodes TCR sequences using ESM and peptides as SMILES strings via MolFormer, capturing both evolutionary and chemical properties. A Multi-Head Cross-Attention (MHCA) module is introduced to align TCR and peptide representations, enabling the model to focus on interaction-relevant features across domains. This architecture improves generalization in zero-shot and few-shot scenarios. Extensive experiments on the TCHard benchmark show that LANTERN consistently outperforms existing baselines including ChemBERTa, TITAN, and NetTCR, particularly on unseen epitope-TCR pairs. These results highlight LANTERN’s potential for robust TCR-pMHC binding prediction and downstream applications in personalized immunotherapy and vaccine development. For reproducing, our code is available at: 
\href{https://anonymous.4open.science/r/LANTERN-87D9/README.md}{https://anonymous.4open.science/r/LANTERN-87D9}.

\end{abstract}

\begin{document}

\flushbottom
\maketitle
\thispagestyle{empty}

\section*{Introduction}

The interaction between T-cell receptors (TCRs) and antigens plays a pivotal role in the adaptive immune response, enabling the immune system to distinguish between self and non-self elements \cite{ref1,ref2}. Accurate modeling of TCR-antigen binding is fundamental to the development of personalized immunotherapies, such as PD-1/PD-L1 blockade, CAR-T therapy, and neoantigen-based vaccines \cite{ref3,ref4}. These treatments depend critically on predicting whether a given TCR sequence will recognize a specific peptide presented by the major histocompatibility complex (pMHC).

Despite extensive progress, computational prediction of TCR-pMHC binding remains an open challenge due to the enormous combinatorial diversity of TCR sequences and the structural complexity of peptide-MHC presentation \cite{ref5,ref6,ref7}. Conventional models often rely solely on sequence similarity, overlooking crucial biochemical and spatial features. Moreover, they tend to perform poorly on unseen or rare TCRs and novel epitopes, limiting their clinical utility \cite{ref8}.

In most recent studies, the complementarity-determining region 3 (CDR3) of the TCR $\beta$ chain has been used as the primary input for modeling, as it is believed to play a dominant role in antigen recognition and is better represented in public datasets \cite{ref9,ref10,ref12,ref13,ref14}. Consistent with this convention, we focus on the CDR3 region of the $\beta$ chain in this study, following prior works such as Weber et al. \cite{ref15} and Lu et al. \cite{ref16}.

Moreover, the reliability of binding prediction depends heavily on the construction of negative samples. While some works utilize biologically verified non-binding pairs from curated databases such as IEDB and NetTCR-2.0 \cite{ref17}, others rely on random pairing strategies \cite{ref17,ref18,ref19} or epitope-frequency-guided sampling. These strategies vary in difficulty and biological realism, and we incorporate multiple such settings in our evaluation to ensure robustness and generalization.

Recent advances in pretraining have also significantly influenced the field. Inspired by self-supervised learning in NLP, models such as TCR-BERT \cite{ref32} have been developed to learn contextual representations from large-scale unlabeled TCR sequences. These models demonstrate improved generalization in downstream binding prediction tasks, particularly in data-scarce settings \cite{ref33,ref34,ref35}. More broadly, the rapid emergence of foundation models in biology has reshaped how complex biological systems are represented and analyzed across scales. Large-scale pretrained models, originally developed for proteins or molecular sequences, are increasingly viewed as foundational components for integrative analysis in single-cell genomics, immune repertoire profiling, and cell–cell interaction modeling~\cite{theodoris2023transfer, cui2024scgpt, qi2025bidirectional}. In single-cell studies, foundation models trained on massive unlabeled corpora enable the extraction of transferable representations that capture functional, evolutionary, and biophysical priors, which are difficult to learn from task-specific datasets alone. These representations have proven especially valuable for downstream tasks under data sparsity, such as rare cell-type identification, immune receptor annotation, and cross-condition generalization.

From this perspective, TCR sequence modeling can be naturally interpreted as a specialized instance of foundation-model-driven single-cell immunology. T-cell receptors arise from clonally expanded single cells, and their sequence diversity and binding specificity reflect both cellular state and immune context. Leveraging pretrained protein language models for TCR representation therefore aligns conceptually with recent efforts to build single-cell foundation models that unify sequence-level information with functional phenotypes. By incorporating large-scale pretrained encoders, our work connects molecular-level immune recognition with the broader paradigm of foundation models for single-cell biology, highlighting how protein language models can serve as modular building blocks for immunological inference within single-cell analysis pipelines.

To address the limitations of existing models, we propose \textbf{LANTERN} (Large lAnguage model-powered TCR-Enhanced Recognition Network), a two-stage deep learning framework that integrates protein language modeling, molecular representation learning, and cross-modality attention to improve generalization in TCR binding prediction. As illustrated in Figure~\ref{fig:framework}, LANTERN adopts a dual-encoder architecture to extract biologically meaningful representations from TCRs and peptides. {\textcolor{black}{Specifically, TCR sequences are processed using ESM, a pretrained protein language model that captures evolutionary and structural information from large-scale protein sequence corpora.
}} In parallel, we convert peptide sequences into SMILES format to represent their chemical graph structures, which are then encoded using MolFormer, a transformer trained for chemical property prediction.

A key innovation of our model lies in how these two representations—TCR (protein) and peptide (molecular)—are fused. Instead of naïvely concatenating their embeddings, we design a \textit{Multi-Head Cross-Attention} (MHCA) module to align the two modalities and allow the TCR embedding to attend selectively to relevant chemical features of the peptide. This mechanism (detailed in Section~\ref{sec:mhca}) enhances the interaction modeling by learning attention patterns across biochemical domains and allows the model to better infer TCR-epitope compatibility.

To train the model, we adopt a binary cross-entropy objective (see Section~\ref{sec:objective}) with optional regularization over the aligned representations. The model is evaluated on multiple reference and random control settings derived from the TCHard dataset, ensuring robust assessment in few-shot and zero-shot scenarios. Our experiments show that LANTERN achieves state-of-the-art performance in predictive accuracy and generalization, particularly when tested on unseen peptides or rare TCRs.

In summary, our contributions are:
\begin{itemize}
    \item We propose LANTERN, a novel TCR-peptide binding prediction framework that leverages protein and molecular pretraining.
    \item We introduce a cross-modality alignment module based on Multi-Head Cross-Attention to model fine-grained biochemical interactions between TCRs and peptides.
    \item We demonstrate that integrating SMILES-based molecular structure improves performance on challenging benchmark splits, particularly in low-resource and zero-shot settings.
\end{itemize}

\section*{Related Work}

\textbf{Modeling TCR-pMHC Interactions.}  
Numerous computational methods have been proposed to model the interaction between TCRs and pMHCs. Early rule-based or similarity-based models such as GLIPH \cite{ref20} and TCRdist \cite{ref11} cluster TCR sequences based on CDR3 region similarity to infer shared specificity. These approaches assume that TCRs with similar sequence motifs tend to recognize the same epitopes. \textcolor{black}{While effective for identifying broad patterns, such models rely on fixed, hand-crafted similarity metrics and are therefore limited in their ability to capture complex, nonlinear interaction signals. In contrast, deep learning–based approaches learn task-specific representations and similarity functions directly from data, enabling more flexible modeling of TCR–peptide interactions.
}

With the growing availability of labeled data, machine learning models have gained popularity. Jurtz et al. \cite{ref24} introduced NetTCR, a 1D convolutional neural network (CNN) that takes both the peptide sequence and the TCR $\beta$ chain CDR3 as input to predict binding. Montemurro et al. \cite{ref25} extended this to NetTCR-2.0 with enhanced architectures for handling variable-length sequences. DeepTCR \cite{ref14} further integrated unsupervised and supervised representation learning to capture sequence motifs across both TCR and antigen spaces. While these models improved predictive accuracy, they typically require large quantities of labeled data.

More recently, transformer-based architectures have been explored. For instance, TEINet \cite{ref18} employs two pretrained encoders—one for TCR and one for epitope sequences—to capture contextual dependencies and enhance generalization. These models reflect a shift from hand-engineered similarity metrics toward end-to-end learnable architectures capable of modeling variable sequence length and interaction context \cite{wang2025modeling}.  However, they often rely on concatenating the two modalities, which limits their ability to model fine-grained alignment between TCR and peptide features.

\textbf{TCR and Peptide Representations.}  
Accurate representation of TCR and peptide sequences is critical to predictive performance. Most studies focus on the CDR3 region of the TCR, particularly the $\beta$ chain, which is known to contribute most significantly to antigen recognition \cite{ref9,ref10,ref12}. The $\alpha$ chain is often omitted due to its relatively minor role and limited public data availability \cite{ref13,ref14}. Peptides are typically short (8–11 amino acids), making it difficult for sequence-based models to capture structural or chemical variability. {\textcolor{black}{To address this, some approaches convert peptide sequences into graph representations or use chemical descriptors such as SMILES strings~\cite{ref15}, which better preserve information about molecular connectivity (bonding) and functional groups, and can, in some cases, encode basic stereochemical information.
}}

\textbf{Cross-Modality Fusion Techniques.}  
A critical design choice in TCR-peptide modeling lies in how the two modalities are combined. The majority of existing models adopt simple concatenation of embeddings, assuming the model will learn relevant interaction signals downstream. However, this approach treats TCR and peptide representations as flat vectors without explicit alignment. Cross-attention mechanisms, such as those used in our LANTERN framework, offer a more principled way to fuse modalities. Inspired by vision-language transformers and recent advances in protein-ligand modeling, cross-attention enables one sequence (e.g., TCR) to selectively attend to relevant regions of the other (e.g., peptide), thereby enhancing the interaction representation. Similar ideas have shown success in protein-protein interaction prediction and drug-target modeling, but remain underexplored in the TCR modeling domain.

\textbf{Pretraining and Self-Supervised Learning.}
Self-supervised learning has emerged as a powerful tool for addressing data scarcity in biological sequence modeling \cite{cui2024scgpt,qi2025bidirectional}. BERT-style models such as TCR-BERT \cite{ref32} and ProGen \cite{ref28} have demonstrated that large-scale masked language modeling can learn transferable representations across tasks. TCR-BERT, in particular, learns semantic embeddings from millions of unlabeled CDR3 sequences, achieving strong generalization on downstream binding prediction \cite{ref33,ref34}. Similarly, ESM \cite{ref30} and ProtTrans \cite{ref31} pretrained on UniRef or BFD datasets have improved performance across multiple protein tasks. In the molecular domain, models like MolBERT and MolFormer leverage graph or SMILES-based pretraining to encode chemical properties. Our approach combines both protein and molecular pretrained models, bridging these two lines of research.

\textbf{Datasets and Negative Sampling.}  
Several curated databases support TCR modeling, including VDJdb \cite{ref21}, IEDB \cite{ref22}, McPAS \cite{ref23}, and the recently curated TCHard dataset \cite{ref17}. These datasets differ in terms of sample size, diversity, and confidence level of binding pairs. Negative sampling remains a critical factor influencing evaluation and training. Biologically validated non-binding pairs offer high-quality negatives but are rare and expensive to obtain. Random pairing strategies are widely used but may introduce false negatives, affecting both training signal and evaluation fairness \cite{ref17,ref18,ref19}. Some studies propose more refined sampling, such as epitope-frequency-guided sampling or structure-informed rejection sampling. In this work, we adopt both reference-based and random negative sampling strategies to thoroughly assess model generalization.

\section*{Methods}
\subsection*{An overview of LANTERN}

\begin{figure}[ht]
\begin{center}
\centerline{\includegraphics[width=\columnwidth]{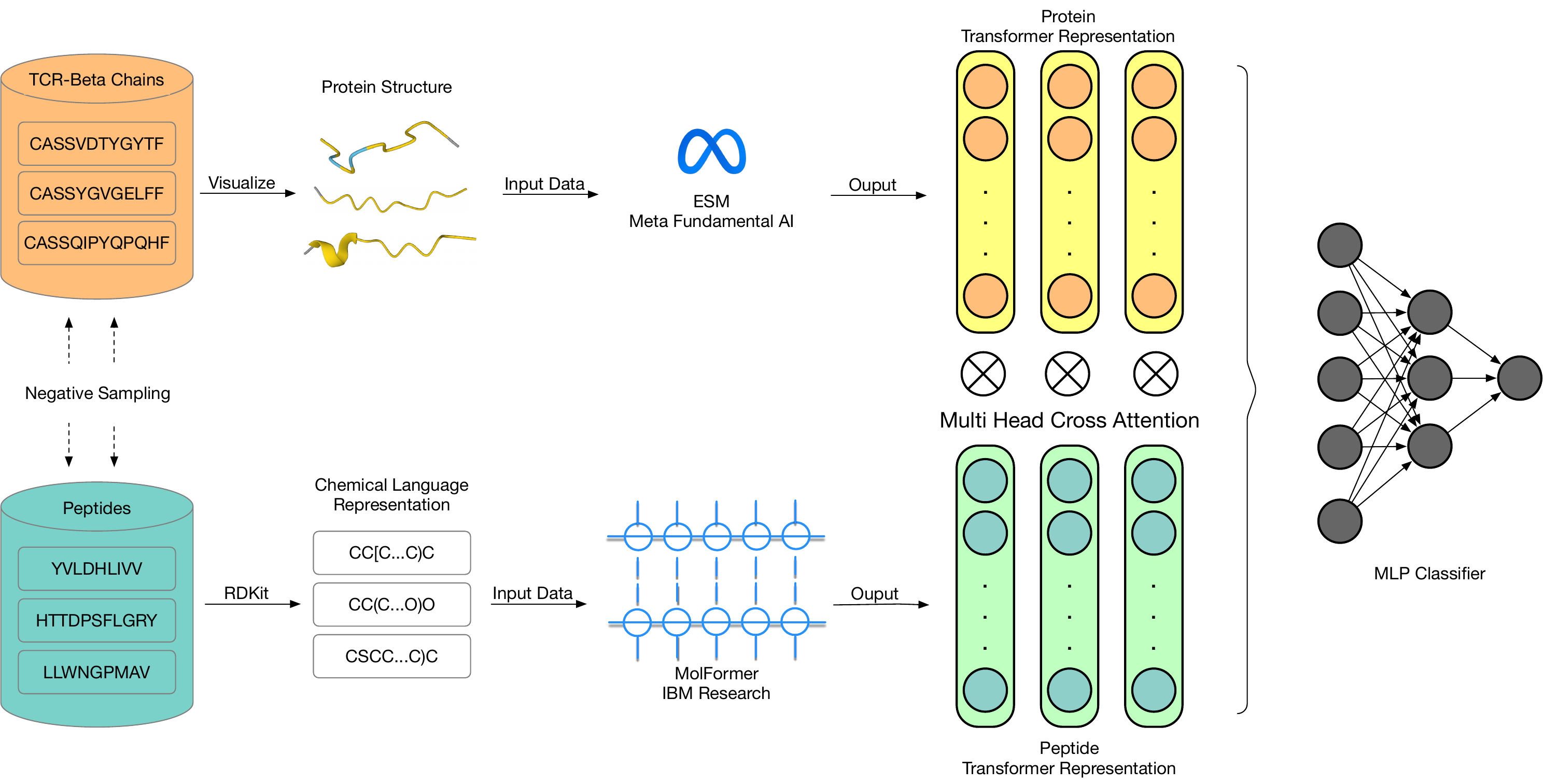}}
\caption{\textbf{Framework of the LANTERN Model.} LANTERN encodes TCR $\beta$ chains using ESM and peptide sequences via SMILES and MolFormer. A Multi-Head Cross-Attention module aligns TCR and peptide embeddings before final classification. The model supports both random and reference-based negative sampling, and generalizes well to unseen epitopes.}
\label{fig:framework}
\end{center}

\end{figure}

In this study, we present the LANTERN model, which employs a two-stage deep learning approach using transfer learning to identify patterns in TCR and epitope sequences. In Figure \ref{fig:framework}, we present a visualization of the LANTERN model. The first stage involves transforming epitope sequences from protein sequences into SMILES sequences to address the issue of short peptide data lengths. {\textcolor{black}{Typically, the epitope peptides are naturally represented as short amino acid sequences, so their limited length (typically 8–11 residues) poses challenges for deep sequence models, as the resulting representations contain limited contextual information and are difficult to pretrain effectively. By converting peptides into SMILES strings, we obtain longer sequences that explicitly encode molecular connectivity and chemical structure, enabling more expressive representations and better compatibility with pretrained molecular language models.}} We begin by tokenizing TCR and epitope sequences into individual characters, which are then processed by their respective pretrained encoders. These encoders convert character sequences into numerical vectors, producing low-dimensional {\textcolor{black}{(256 dimensions in our implementation)}} vector representations. Finally, the encodings of TCRs and epitopes are integrated and fed into a fully connected neural network designed to leverage information from both parts to make accurate predictions.

\subsection*{Negative Sampling Strategy}

In the context of the TCR-peptide binding problem, two primary strategies are commonly employed for generating negative samples: the reference control and the random control \cite{ref17}.

\textbf{Reference Control}: This approach involves negative samples derived from established datasets, providing verified unbound pairs confirmed through biological experiments. While this method offers reliable ground truth data, it is resource-intensive and requires substantial time and experimental effort, posing challenges for large-scale datasets.

\textbf{Random Control}: This method generates negative samples by randomly pairing a peptide with a TCR sequence. Although this approach is efficient and requires minimal additional resources, it carries the risk of producing pairs that may still bind, potentially compromising the accuracy of the negative sample set.

Among these strategies, the random control poses a greater challenge due to the possibility of inadvertently generating binding pairs. To ensure the model is evaluated on unseen peptides during testing (i.e., peptides in the test set are absent from the training data), our negative sampling procedure constructs an equal number of negative pairs to match the positive ones in the training set. Specifically, we generate negative samples by randomly pairing one peptide with one TCR from the training data, maintaining the data distribution consistent with the TCHard dataset. Through this process, we generate the corresponding reference control dataset, denoted as Gen NA, and the corresponding random control dataset, denoted as Gen RN. This procedure is formally defined by the following equations:


\[
D_{\text{neg}} = 
\left\{
(TCR_i, Pep_j) \mid 
\begin{array}{l}
    TCR_i \in TCR_{\text{train}}, \\
    Pep_j \in Pep_{\text{train}}, \\
    (TCR_i, Pep_j) \notin D_{\text{pos}}
\end{array}
\right\}
\]

\[
\text{such that} \quad |D_{\text{neg}}| \propto |D_{\text{pos}}|
\]

where
\begin{itemize}
    \item $TCR_{train}$ is the set of TCR sequences in the training set.
    \item $Pep_{train}$ is the set of peptide sequences in the training set.
    \item $D_{neg}$ is the set of generated negative samples.
    \item $D_{pos}$ is the set of positive samples in the training set.
\end{itemize}

\subsection*{Encoder of TCR and Peptides}
\label{sec:mhca}
We first encode the input TCR (protein) and peptide (molecular) sequences into vector representations using domain-specific pretrained encoders: ESM for TCR and MolFormer for peptide SMILES.

\subsubsection*{TCR Encoding via ESM}

Given a TCR amino acid sequence $S_{\text{protein}} = \{a_1, a_2, ..., a_L\}$, we tokenize it into individual characters, then embed each into an input matrix $X \in \mathbb{R}^{L \times d}$.

We first apply Layer Normalization:

\begin{equation}
\centering
\widetilde{X} = \text{LN}(X)
\end{equation}

Then we apply Multi-Head Self-Attention (MHSA):

\begin{equation}
\centering
\text{MHSA}(\widetilde{X}) = \text{Concat}(head_1, head_2, ..., head_h)W^O
\end{equation}

Each attention head models intra-TCR amino acid interactions and is defined as:

\begin{equation}
head_i = \text{Softmax}\left( \frac{Q_i K_i^T}{\sqrt{d_k}} \right) V_i
\end{equation}

where the queries, keys, and values are projections of the normalized input:

\begin{equation}
Q_i = \widetilde{X} W_i^Q, \quad K_i = \widetilde{X} W_i^K, \quad V_i = \widetilde{X} W_i^V
\end{equation}

After aggregating local/global dependencies among TCR residues, we apply a Feed-Forward Network:

\begin{equation}
E_{\text{protein}} = \text{FFN}\left(\widetilde{X} + \text{MHSA}(\widetilde{X})\right)
\end{equation}

\begin{equation}
\text{FFN}(x) = \text{ReLU}(xW_1 + b_1)W_2 + b_2
\end{equation}

where $W_1$, $W_2$, $b_1$, $b_2$ are learnable parameters. The resulting $E_{\text{protein}}$ summarizes the functional representation of the TCR sequence, encoding biochemical and spatial relationships relevant to antigen recognition.

\subsubsection*{Peptide Encoding via MolFormer}

The peptide (epitope) sequence $P_{\text{peptide}}$ is first converted to a SMILES string using depth-first traversal to simulate its molecular graph:

\begin{equation}
P_{\text{molecule}} = \text{DFS}(P_{\text{peptide}})
\end{equation}

This sequence is then tokenized and embedded into $Y \in \mathbb{R}^{T \times d}$ for transformer processing. As with ESM, we apply Layer Normalization and MHSA:

\begin{equation}
\widetilde{Y} = \text{LN}(Y)
\end{equation}

\begin{equation}
\text{MHSA}(\widetilde{Y}) = \text{Concat}(head_1', ..., head_h')W^{O'}
\end{equation}

where each head captures intra-molecular interactions within the peptide:

\begin{equation}
head_j' = \text{Softmax}\left( \frac{Q_j' {K_j'}^T}{\sqrt{d_k}} \right) V_j'
\end{equation}

\begin{equation}
Q_j' = \widetilde{Y} W_j^{Q'}, \quad K_j' = \widetilde{Y} W_j^{K'}, \quad V_j' = \widetilde{Y} W_j^{V'}
\end{equation}

The molecular embedding is produced by:

\begin{equation}
E_{\text{molecule}} = \text{FFN}\left(\widetilde{Y} + \text{MHSA}(\widetilde{Y})\right)
\end{equation}

\subsubsection*{Cross-Modality Alignment via MHCA}

To align TCR and peptide representations, we apply a Multi-Head Cross-Attention mechanism where $E_{\text{protein}}$ serves as query and $E_{\text{molecule}}$ provides key-value context, where $[\cdot ; \cdot]$ denotes vector concatenation.:

\begin{equation}
Q = E_{\text{protein}} W^Q, \quad K = E_{\text{molecule}} W^K, \quad V = E_{\text{molecule}} W^V
\end{equation}

\begin{equation}
head_i = \text{Softmax}\left( \frac{Q_i K_i^T}{\sqrt{d_k}} \right) V_i, \quad \text{for } i = 1, ..., h
\end{equation}

\begin{equation}
MHCA(Q,K,V) = [\text{head}_1; \ldots; \text{head}_h] W^O
\end{equation}

This results in an aligned representation:

\begin{equation}
E_{\text{align}} = \text{MHCA}[E_{\text{protein}}; E_{\text{molecule}}]
\end{equation}

\subsubsection*{Binding Affinity Prediction}

We concatenate the aligned representation with the original TCR embedding for final classification:

\begin{equation}
Z = [E_{\text{protein}}; E_{\text{align}}]
\end{equation}

\begin{equation}
\hat{y} = \sigma(W_{\text{MLP}} Z + b_{\text{MLP}})
\end{equation}

where $\hat{y}$ is the predicted binding affinity score (0–1), and $\sigma$ denotes the sigmoid activation for binary classification.

\subsection*{Objective Function}
\label{sec:objective}

The primary task of our model is to predict the binary binding outcome between a TCR sequence and a peptide (encoded as a SMILES molecule). The output of the model is a scalar $\hat{y} \in (0,1)$ representing the probability that a given TCR binds to a given peptide.

We formulate this as a binary classification task and use the Binary Cross-Entropy (BCE) loss as the main training objective:

\begin{equation}
\mathcal{L}_{\text{BCE}} = - \frac{1}{N} \sum_{i=1}^{N} \left[ y_i \log(\hat{y}_i) + (1 - y_i) \log(1 - \hat{y}_i) \right]
\end{equation}

where:
\begin{itemize}
    \item $y_i \in \{0,1\}$ is the ground-truth label for the $i$-th TCR-peptide pair ($1$ for binding, $0$ for non-binding),
    \item $\hat{y}_i$ is the predicted binding probability from the final MLP layer,
    \item $N$ is the number of training examples.
\end{itemize}

\subsubsection*{Cross-Attention Learning:}

The Multi-Head Cross-Attention (MHCA) module is trained jointly with the prediction head. Its learnable parameters $\{W^Q, W^K, W^V, W^O\}$ are updated via backpropagation through the BCE loss $\mathcal{L}_{\text{BCE}}$, enabling the model to learn optimal alignment between $E_{\text{protein}}$ (TCR) and $E_{\text{molecule}}$ (peptide) representations.

The gradients of $\mathcal{L}_{\text{BCE}}$ with respect to the model parameters are computed via backpropagation, including the parameters of the MHCA module. Formally, for each head $i$, the gradients are computed by:

\begin{equation}
\frac{\partial \mathcal{L}_{\text{BCE}}}{\partial W_i^Q},\quad
\frac{\partial \mathcal{L}_{\text{BCE}}}{\partial W_i^K},\quad
\frac{\partial \mathcal{L}_{\text{BCE}}}{\partial W_i^V},\quad
\frac{\partial \mathcal{L}_{\text{BCE}}}{\partial W^O}
\end{equation}

{\textcolor{black}{This design encourages the model to attend to interaction-relevant features (e.g., CDR3 motifs and chemical substructures), facilitating more structured cross-modality alignment and improved binding prediction performance.}}

\subsubsection*{Alignment Regularization:}

An auxiliary regularization term is added to further constrain the alignment between $E_{\text{protein}}$ and $E_{\text{align}}$. {\textcolor{black}{The alignment regularization term encourages consistency between the original TCR representation $E_{\text{protein}}$ and the cross-attention–enhanced representation $E_{\text{align}}$. Since $E_{\text{align}}$ is computed using $E_{\text{protein}}$ as the query in the MHCA module, this regularization stabilizes the TCR-centric interaction representation while allowing peptide-specific modulation.}} For instance, we encourage similarity via mean squared error:

\begin{equation}
\mathcal{L}_{\text{align}} = \frac{1}{N} \sum_{i=1}^{N} \| E_{\text{align}}^{(i)} - E_{\text{protein}}^{(i)} \|^2
\end{equation}

The final loss function is then:

\begin{equation}
\mathcal{L} = \mathcal{L}_{\text{BCE}} + \lambda \cdot \mathcal{L}_{\text{align}}
\end{equation}

where $\lambda$ is a tunable hyperparameter controlling the weight of alignment regularization.

\section*{Results}

To evaluate the effectiveness of LANTERN in TCR-peptide binding prediction, we conduct a series of experiments comparing it with state-of-the-art baselines across multiple datasets and settings. Our experiments assess both zero-shot and few-shot learning capabilities, as well as the quality of learned embeddings. The code for reproducing our experiments are available at: 

\href{https://anonymous.4open.science/r/LANTERN-87D9/README.md}{https://anonymous.4open.science/r/LANTERN-87D9}.

\subsection*{Dataset Statistics}

We prepare data from the TCHard dataset \cite{ref17} according to the negative sampling strategy. This results in the original reference control dataset (NA), the original random control dataset (RN), the generated reference control dataset (Gen NA), and the generated random control dataset (Gen RN). For all four datasets, we select the TCR Beta chain and the antigen peptide as the input feature, ignoring the TCR Alpha chain, MHC, V, and J genes. This selection is based on findings from the reference paper \cite{ref36,ref37,ref38}, which indicates that the Beta chain and antigen peptide are the most critical components for predicting TCR-peptide interactions. By focusing on these elements, we aim to improve prediction accuracy while simplifying the model by excluding less impactful features. We employ a 5-fold split strategy, ensuring that in each split, the original test set from TCHard is used consistently across experiments, and for the original training set, we further split it into a new training set and a validation set in an 8:2 ratio. Throughout all datasets, we maintain a "hard split" methodology, ensuring that peptides in the test or validation set are never seen in the training set. Totally, there are ~160,000 TCR beta chains, and ~1,000 peptides, resulting in ~220,000 TCR-Peptide pairs in the train set, ~60,000 validation set and ~40,000 TCR-Peptide pairs in the test set. Below we show the statistics of the reference control and random control settings

\begin{table}[ht]
\centering
\caption*{\textbf{Table 1.} Statistics of the 5-fold cross validation of reference control setting}
\begin{tabular}{|c|c|c|c|c|c|}
\hline
& \# pairs (Train) & \# pairs (Val) & \# pairs (Test) & \# unique pep (Train) & \# unique pep (Val) \\
\hline
CV1 & 228481 & 62859 & 40480 & 816 & 525 \\
CV2 & 225668 & 61838 & 43293 & 813 & 521 \\
CV3 & 228411 & 62500 & 40550 & 813 & 521 \\
CV4 & 228515 & 62210 & 40446 & 816 & 521 \\
CV5 & 228550 & 62775 & 40411 & 815 & 524 \\
\hline
\end{tabular}
\end{table}

\begin{table}[ht]
\centering
\caption*{\textbf{Table 2.} Statistics of the 5-fold cross validation of random control setting}
\begin{tabular}{|c|c|c|c|c|c|}
\hline
& \# pairs (Train) & \# pairs (Val) & \# pairs (Test) & \# unique pep (Train) & \# unique pep (Val) \\
\hline
CV1 & 340674 & 82674 & 60629 & 860 & 468 \\
CV2 & 340549 & 68396 & 60754 & 874 & 458 \\
CV3 & 338122 & 69958 & 63181 & 877 & 459 \\
CV4 & 340808 & 70030 & 60495 & 876 & 459 \\
CV5 & 340542 & 68507 & 60761 & 874 & 459 \\
\hline
\end{tabular}
\end{table}

\subsection*{Benchmark Evaluation of LANTERN}

The objective is to predict whether the TCR and peptide bind with each other. In LANTERN, we choose MolFormer as the peptide SMILES pretraining model and use other state-of-the-art models as baselines for performance comparison. These include Large Language Model-based approaches, including Smiles-Bert \cite{ref39}, TEINet \cite{ref18}, and ChemBERTa \cite{ref40}, as well as end-to-end deep learning models such as TITAN \cite{ref15}, ERGO II \cite{ref7}, NetTCR \cite{ref25}, Dlptcr \cite{ref41}, and Imrex \cite{ref42}. The main results and analysis demonstrate the powerful performance of Large Language Models in zero-shot learning, effectively capturing the interaction between TCR and peptide. Additionally, their few-shot learning capabilities highlight their ability to remember and generalize knowledge from other domains to the binding prediction problem.

The model is trained end-to-end using the Adam optimizer with early stopping based on validation AUC. All pretrained encoders (ESM and MolFormer) are frozen or fine-tuned depending on experimental settings. The MHCA module and MLP layers are trained from scratch. The model's performance is rigorously validated against a held-out test set, employing metrics such as AUC-ROC to quantify prediction accuracy and reliability. This methodology of LANTERN, integrating the powerful ESM-1b and MolFormer models, sets the stage for profound advancements in understanding and predicting the dynamics of protein-ligand interactions, with significant implications for drug discovery and biochemical research.

\subsection*{Performance of LANTERN Model}
\begin{figure}[ht]
\begin{center}
\centerline{\includegraphics[width=\columnwidth]{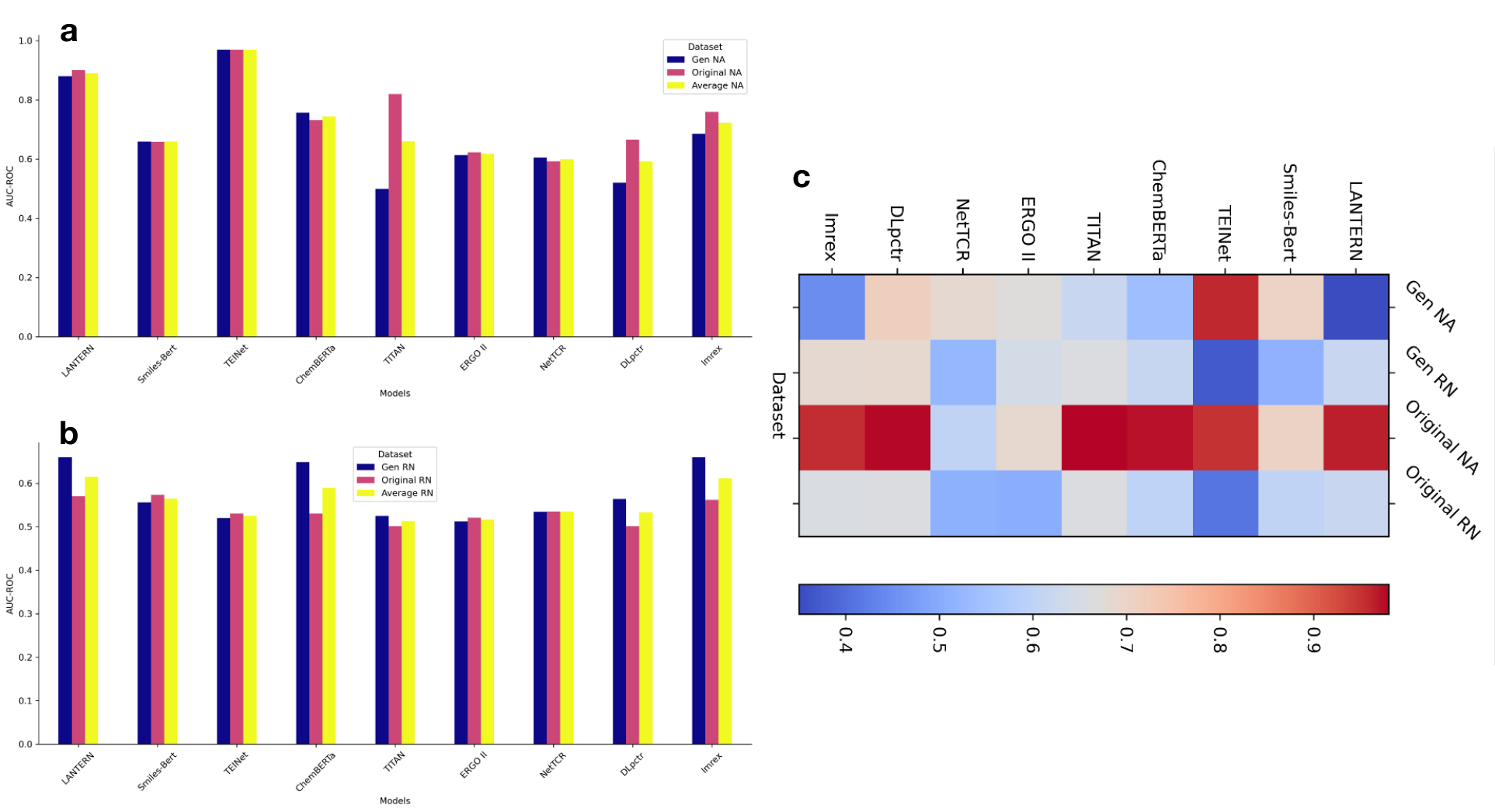}}
\caption{\textbf{Performance of LANTERN on four datasets. |} \textbf{a.} AUC-ROC scores across models for NA dataset. \textbf{b.} AUC-ROC scores across models for RN dataset. \textbf{c.} Accuracy scores across models for the four datasets.}
\label{fig:figure2}
\end{center}
\end{figure}

LANTERN exhibits significant improvement over traditional methods. Our experiments highlighted the performance of the LANTERN framework, particularly in the context of the random control setting. Figure \ref{fig:figure2} presents the AUC-ROC score across the four settings, for the detailed results of all metrics, please refer to the Table 3 to Table 6. We use six metrics to evaluate performance: AUC-ROC, Accuracy, Recall, Precision, F1-Score, and AUC-PR. Among these, we primarily focus on AUC-ROC, as it is robust to threshold values in binary classification tasks.

\begin{table}[ht]
\centering
\caption*{\textbf{Table 3.} Result of LANTERN and baseline models on Gen NA}
\begin{tabular}{|l|c|c|c|c|c|c|}
\hline
\textbf{Model} & \textbf{AUC-ROC} & \textbf{ACC} & \textbf{RECALL} & \textbf{PRECISION} & \textbf{F1-SCORE} & \textbf{AUC-PR} \\
\hline
LANTERN & 0.882 & 0.728 & 0.341 & 0.992 & 0.483 & 0.991 \\
Smiles-Bert & 0.659 & 0.702 & 0.621 & 0.843 & 0.811 & 0.808 \\
TEINet & 0.970 & 0.960 & 0.980 & 0.980 & 0.980 & 0.940 \\
ChemBERTa & 0.757 & 0.540 & 0.529 & 0.987 & 0.661 & 0.986 \\
TITAN & 0.502 & 0.620 & 0.523 & 0.590 & 0.608 & 0.970 \\
ERGO II & 0.613 & 0.665 & 0.618 & 0.873 & 0.702 & 0.698 \\
NetTCR & 0.605 & 0.683 & 0.592 & 0.793 & 0.721 & 0.688 \\
DlpTcr & 0.520 & 0.716 & 0.307 & 0.017 & 0.030 & 0.514 \\
Imrex & 0.685 & 0.448 & 0.433 & 0.985 & 0.593 & 0.986 \\
\hline
\end{tabular}
\end{table}

\begin{table}[ht]
\centering
\caption*{\textbf{Table 4.} Result of LANTERN and baseline models on Gen RN}
\begin{tabular}{|l|c|c|c|c|c|c|}
\hline
\textbf{Model} & \textbf{AUC-ROC} & \textbf{ACC} & \textbf{RECALL} & \textbf{PRECISION} & \textbf{F1-SCORE} & \textbf{AUC-PR} \\
\hline
LANTERN & 0.662 & 0.620 & 0.551 & 0.460 & 0.503 & 0.537 \\
Smiles-Bert & 0.556 & 0.513 & 0.498 & 0.443 & 0.468 & 0.491 \\
TEINet & 0.520 & 0.370 & 0.980 & 0.350 & 0.510 & 0.470 \\
ChemBERTa & 0.649 & 0.614 & 0.542 & 0.446 & 0.488 & 0.511 \\
TITAN & 0.525 & 0.660 & 0.537 & 0.516 & 0.523 & 0.551 \\
ERGO II & 0.512 & 0.643 & 0.503 & 0.516 & 0.534 & 0.503 \\
NetTCR & 0.534 & 0.524 & 0.557 & 0.563 & 0.609 & 0.524 \\
DlpTcr & 0.564 & 0.688 & 0.951 & 0.692 & 0.801 & 0.838 \\
Imrex & 0.660 & 0.685 & 0.113 & 0.731 & 0.194 & 0.518 \\
\hline
\end{tabular}
\end{table}

\begin{table}[ht]
\centering
\caption*{\textbf{Table 5.} Result of LANTERN and baseline models on Original NA}
\begin{tabular}{|l|c|c|c|c|c|c|}
\hline
\textbf{Model} & \textbf{AUC-ROC} & \textbf{ACC} & \textbf{RECALL} & \textbf{PRECISION} & \textbf{F1-SCORE} & \textbf{AUC-PR} \\
\hline
LANTERN & 0.901 & 0.964 & 0.997 & 0.966 & 0.981 & 0.995 \\
Smiles-Bert & 0.658 & 0.705 & 0.637 & 0.890 & 0.821 & 0.928 \\
TEINet & 0.972 & 0.951 & 0.991 & 0.963 & 0.981 & 1.000 \\
ChemBERTa & 0.731 & 0.972 & 0.994 & 0.977 & 0.986 & 0.993 \\
TITAN & 0.820 & 0.980 & 0.999 & 0.990 & 0.990 & 1.000 \\
ERGO II & 0.623 & 0.689 & 0.621 & 0.857 & 0.603 & 0.725 \\
NetTCR & 0.593 & 0.602 & 0.615 & 0.813 & 0.498 & 0.632 \\
DlpTcr & 0.666 & 0.977 & 0.336 & 0.553 & 0.389 & 0.474 \\
Imrex & 0.760 & 0.954 & 0.987 & 0.965 & 0.976 & 0.988 \\
\hline
\end{tabular}
\end{table}

\begin{table}[ht]
\centering
\caption*{\textbf{Table 6.} Result of LANTERN and baseline models on Original RN}
\begin{tabular}{|l|c|c|c|c|c|c|}
\hline
\textbf{Model} & \textbf{AUC-ROC} & \textbf{ACC} & \textbf{RECALL} & \textbf{PRECISION} & \textbf{F1-SCORE} & \textbf{AUC-PR} \\
\hline
LANTERN & 0.570 & 0.620 & 0.260 & 0.400 & 0.320 & 0.400 \\
Smiles-Bert & 0.573 & 0.602 & 0.521 & 0.763 & 0.702 & 0.817 \\
TEINet & 0.531 & 0.417 & 0.880 & 0.351 & 0.504 & 0.353 \\
ChemBERTa & 0.530 & 0.601 & 0.213 & 0.339 & 0.226 & 0.352 \\
TITAN & 0.501 & 0.660 & 0.413 & 0.562 & 0.431 & 0.347 \\
ERGO II & 0.521 & 0.505 & 0.537 & 0.689 & 0.531 & 0.625 \\
NetTCR & 0.535 & 0.511 & 0.547 & 0.621 & 0.509 & 0.662 \\
DlpTcr & 0.501 & 0.658 & 0.991 & 0.660 & 0.792 & 0.829 \\
Imrex & 0.562 & 0.657 & 0.026 & 0.387 & 0.046 & 0.377 \\
\hline
\end{tabular}
\end{table}

Additionally, the LANTERN model demonstrates robust performance, significantly outperforming baseline methods, which are reflected in the following three key observations. Specifically, Large Language Models (LLMs) consistently outperforms end-to-end deep learning models. The mean AUC-ROC score for LLM-based models of 0.8165, compared to 0.5846 for deep learning models, showing an average increase of 0.2319 for LLM-based models. Specifically, LANTERN achieved an AUC-ROC score of 0.88 on Gen NA, significantly higher than many other models. This demonstrates the superior capability of LLM-based models in capturing complex interactions between TCR and peptides. The success of LLM-based models can be attributed to their ability to leverage extensive pretraining on diverse datasets, enabling them to generalize effectively to new, unseen data. This zero-shot learning capability allows these models to recognize patterns and relationships that end-to-end models may miss without extensive retraining.

\subsubsection*{Comparison of Generated vs. Original Datasets}

Comparing results between Gen NA and Original NA, and Gen RN and Original RN reveal that models trained on generated datasets generally performed better than those trained on original datasets. For instance, LANTERN achieves an AUC-ROC score of 0.88 on Gen NA versus 0.901 on Original NA, and 0.66 on Gen RN versus 0.57 on Original RN. This suggests that the novel negative sample generation method improved model training by providing more realistic and challenging scenarios, closely mimicking the complexities of real-world data. The generation of negative samples in Gen NA and Gen RN datasets allowed for a more balanced and comprehensive training process. This method ensures that the model encounters a variety of binding and non-binding interactions, enhancing its ability to distinguish between them. By incorporating a diverse range of negative samples, the model becomes more adept at recognizing subtle differences in binding patterns, leading to improved prediction accuracy and robustness.

\subsubsection*{Generally Challengeable Random Control Setting}

Although LANTERN showed higher performance over the random control setting, there is still room for improvement, as the AUC-ROC score of 0.66 is relatively low. The random control setting is more challenging than the reference control setting due to its nature of generating negative samples quickly and resource-efficiently. However, it requires further exploration to enhance robustness and address this challenge effectively. The random control setting presents unique challenges due to the inherent variability and randomness of the generated negative samples. While this approach is efficient and scalable, it can introduce noise and ambiguity into the training process. To address these challenges, future research could explore advanced techniques for negative sample generation, such as leveraging domain-specific knowledge or incorporating additional biological context to improve the quality and relevance of the negative samples. Additionally, integrating ensemble learning methods or incorporating multimodal data sources could further enhance the model's performance and generalization capabilities.

\subsubsection*{Embedding Cluster of Pretraining Model}
\begin{figure}[ht]
\begin{center}
\centerline{\includegraphics[width=\columnwidth]{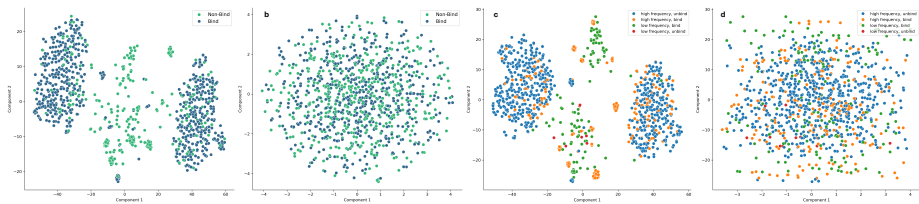}}
\caption{\textbf{t-SNE visualization of the embeddings from LANTERN model. |} \textbf{a.} With pretrain of 0.857 h score. \textbf{b.} Without pretrain of 0.607 h score. \textbf{c.} Show frequency of the pretrained embeddings. \textbf{d.} Show frequency of the no pretrained embeddings.}
\label{fig:figure3}
\end{center}
\end{figure}

To visualize the impact of using pretraining models on our embeddings, we clustered the TCR and antigen embeddings generated by LANTERN with and without the utilization of ESM and MolFormer pretraining. The Hopkins score for the pretraining model is 0.90, and the Cluster Tendency Score (CTS) is 0.86. Specifically, we sampled 10,000 embeddings from the pretrained model and reset the labels to the ground truth of binding or unbinding. Using the t-SNE method to reduce the dimensions to a 2D plane, we obtained the clusters shown in Figure 3.

From Figure 3a and 3b, it is evident that there is a clear boundary between the two classes (bind and non-bind), making it easier for the classifier to distinguish between them. This separation illustrates the effectiveness of the pretraining in enhancing the clustering of embeddings, thus aiding the model’s classification performance.
Additionally, we evaluated whether the pretrained embeddings could reflect the property of low and high frequency. We mapped the low/high frequency to the first dimension, resulting in four classes: [low frequency, unbind], [low frequency, bind], [high frequency, unbind], and [high frequency, bind], as shown in Figure 3c and 3d. The figures show that the model can also distinguish between high and low frequencies effectively.
These visualizations demonstrate that the pretrained embeddings not only provide clear separations between binding and non-binding classes but also accurately reflect the frequency characteristics, further supporting the robustness and effectiveness of the pretraining approach in our model.

\subsubsection*{Quantitative Evaluation of Embedding Performance}

\subsubsection*{Embedding Cluster of Pretraining Model}
\begin{figure}[ht]
\begin{center}
\centerline{\includegraphics[width=\columnwidth]{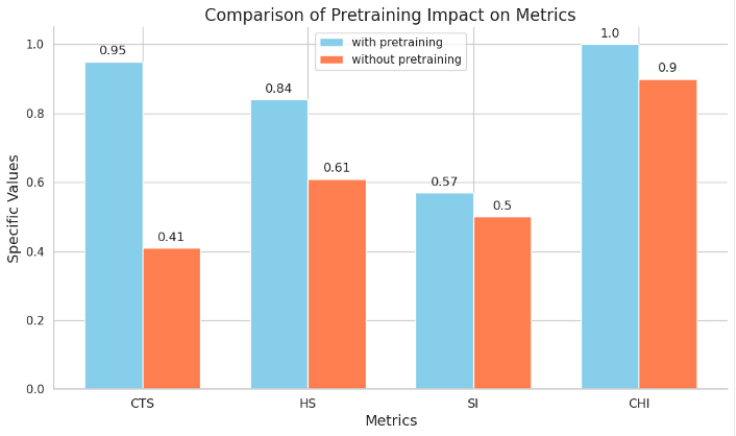}}
\caption{Comparison of cluster metrics between embeddings with or without pretraining. The SI and CHI scores are scaled by sigmoid function to make them more readable.}
\label{fig:figure4}
\end{center}
\end{figure}

To further quantify the performance of the embeddings learned by the pretraining model, we calculated several clustering performance metrics: Silhouette Index (SI) and Calinski-Harabasz Index (CHI). Additionally, we evaluated unsupervised metrics, including the Hopkins Score (HS) and Cluster Tendency Score (CTS) \cite{ref43}. For SI and CH, we employed the K-means clustering method on the embeddings with  k = 2. We put the algorithm for these metrics in the supplementary section 1. The results are illustrated in the Figure \ref{fig:figure4}.
In Figure 4, we observe the following metrics:

• Cluster Tendency Score (CTS): With pretraining, CTS achieved a score of 0.96, significantly higher than the 0.41 score without pretraining. This indicates that the pretraining model substantially improves the clustering tendency of the embeddings.

• Hopkins Score (HS): The Hopkins Score with pretraining is 0.84, compared to 0.61 without pretraining. A higher Hopkins Score suggests better cluster formation, reinforcing the effectiveness of pretraining.
    
• Silhouette Index (SI): The SI score is 0.31 with pretraining and 0.00 without pretraining, demonstrating better-defined clusters with pretraining.
    
• Calinski-Harabasz Index (CHI): The CHI with pretraining is 1. (scaled from 389.06), dramatically higher than the 0.9 (scaled from 2.12) without pretraining, indicating that the embeddings with pretraining result in more compact and well-separated clusters.

These metrics collectively illustrate the advantage of using LANTERN with pretraining. The significant improvements in CTS, HS, SI, and CHI scores indicate that the pretraining enhances the quality of the embeddings, leading to better-defined and more distinguishable clusters. This improved clustering performance directly contributes to the model’s ability to accurately predict TCR-peptide interactions, showcasing the robustness and effectiveness of the pretraining approach.

\subsection*{Few-Shot Learning Capabilities of LANTERN}

\begin{figure}[ht]
\begin{center}
\centerline{\includegraphics[width=\columnwidth]{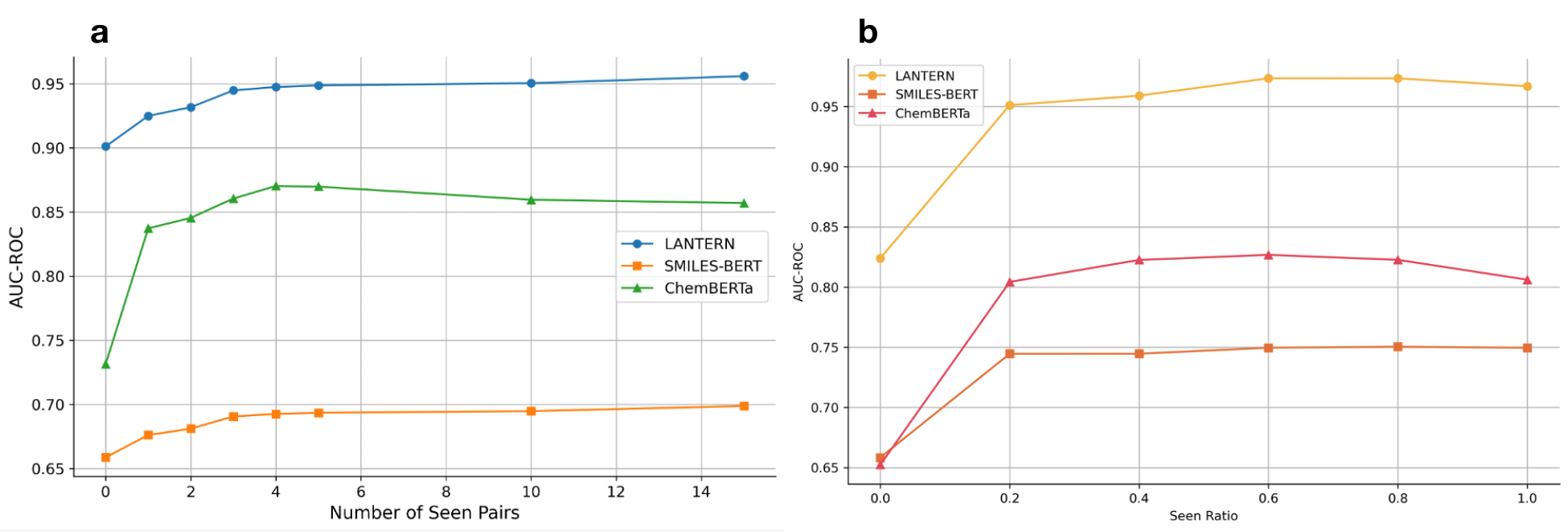}}
\caption{\textbf{Few-Shot Learning performance of LANTERN |} a. Performance comparison of models over different numbers of seen pairs. b. Performance comparison of models over different seen ratios.}
\label{fig:figure5}
\end{center}
\end{figure}

Previous sections highlighted the powerful performance of LANTERN in zero-shot learning settings. Next, we explore the model’s performance in few-shot learning scenarios. The few-shot learning capabilities of LLM-based models enable them to adapt quickly to new tasks with minimal additional training data, demonstrating their flexibility and robustness in various contexts. Few-shot learning is a more common setting encountered in practice. It involves training the model with only a limited number of samples (5 or fewer) of TCR and peptide pairs, aiming to achieve better performance from these limited samples. If the model can learn effectively from a few samples, it becomes significantly more useful in real-world applications.
To introduce the few-shot learning setting, we control two variables: the number of peptides seen by the training set and the number of TCR-peptide pairs for each seen peptide in the training set. By carefully managing these variables, we can evaluate LANTERN’s ability to generalize and perform well with minimal data, which is crucial for practical applications where data is often scarce.

In Figure 5a, we maintain 20\% seen ratio for the peptide in the training set and test set, then we gradually increase the number of TCR-Peptide pairs for each peptide in the training set. The result suggests two points: 1. the performance of LANTERN and baseline models shows a rapid increase with only a few peptide-TCR pairs (fewer than 5) in the training set. This is particularly beneficial as it indicates that even with limited experimental data, the model can effectively understand the binding or non-binding information of the peptide, reducing the need for extensive data. 2. The performance increase speed is gradually slow when the TCR-Peptide pairs increase, this trend suggests that the model reaches saturation to the seen data, if we want to further improve the performance, we should add more and more data to the model. This also indirectly supports the scaling law.

In Figure 5b, we control the number of TCR-Peptide pairs, limiting it to 5 pairs for each peptide in the training set, and gradually increase the ratio of seen peptides in the training set. The results indicate that as the seen ratio increases, the performance of the models improves steadily. Specifically, the MoleFormer model shows a steep increase in AUC-ROC as the seen ratio rises, reaching near-perfect accuracy when the seen ratio is at its maximum. This trend suggests that MoleFormer is highly effective at predicting with increasing exposure to seen peptides, achieving almost 100\% accuracy under a fully seen setting. In contrast, SMILES-BERT and ChemBERTa demonstrate more modest improvements, with SMILES-BERT showing the least sensitivity to the increasing seen ratio, plateauing at a lower AUC-ROC compared to the other models.

\section*{Discussion and Conclusion}

LANTERN advances TCR-pMHC binding prediction by combining protein and molecular language models with cross-attention-based fusion. Its use of ESM and MolFormer embeddings enables the model to capture both sequence-level and chemical features, leading to strong performance in zero-shot and few-shot settings—particularly when predicting unseen epitopes. A key strength of LANTERN lies in its ability to generalize from limited data. Across various negative sampling strategies, the model consistently outperforms baseline methods, including both traditional deep learning and recent LLM-based approaches. The Multi-Head Cross-Attention module enhances interpretability and improves alignment between TCR and peptide features. While LANTERN performs well in both reference and random control settings, there is still room for improvement in distinguishing hard negatives. Future work may explore improved negative sampling, structural priors, or multimodal integration to further boost performance.

In conclusion, LANTERN represents a notable step forward in the field of TCR-pMHC binding prediction. Its ability to generalize across novel epitopes and its strong performance in low-data environments make it a valuable tool for advancing personalized immunotherapies and vaccines. Future research will likely focus on enhancing its robustness and exploring additional biological contexts to push the boundaries of what is possible in TCR-pMHC interaction prediction.

\section*{Acknowledgments}

So long and thanks for all the fish.

\bibliography{sample}

\appendix
\onecolumn

\section*{Evaluation Metrics}

\subsection*{Hopkins Score}

The Hopkins Score (HS) is a statistical test that measures the clustering tendency of a dataset, helping to determine whether the data points are more likely to form clusters or are uniformly distributed. The method begins by selecting a random subset of data points from the dataset. For each of these data points, a corresponding random point is generated within the same range as the data points. The next step involves calculating the distances between the nearest neighbors: $D_T$ represents the sum of the distances from each selected data point to its nearest neighbor within the subset, while $D_R$ represents the sum of the distances from each random point to its nearest data point. The Hopkins Score is then calculated using the formula $HS = \frac{D_R}{D_R + D_T}$. A Hopkins Score close to 1 indicates that the data has a strong clustering tendency, whereas a score around 0.5 suggests that the data points are uniformly distributed, showing no significant clustering.

\subsection*{Cluster Tendency Score}

The Cluster Tendency Score (CTS)~\cite{ref43} assesses how well region embeddings form clusters, under the assumption that clustered embeddings are more useful than dispersed ones. Inspired by the Hopkins test, the process starts with a set of $N_S = \{q_i \mid i = 1, ..., N_S\}$ embeddings. From $N_S$, a set of $N_T$ embeddings is selected as test points, denoted as $T = \{p_i \mid i = 1, ..., N_T\}$. For each test point $p_i$, its nearest neighbor $\hat{q}_i$ is found within $N_S \setminus \{q_i\}$. The sum of squared Euclidean distances between each test point $p_i$ and its nearest neighbor is calculated as $D_T = \sum_i = 1^{N_T} d(p_i, \hat{q}_i).$ Next, the same number of $N_T$ random points $U = \{u_i \mid i = 1, ..., N_T\}$ are sampled uniformly between the minimum and maximum values across each dimension of the embeddings in $N_S$. The sum of distances between each random point $u_i$ and its nearest neighbor $\hat{q}_i$ is calculated as $D_R = \sum_i = 1^{N_T} d(u_i, \hat{q}_i).$ The CTS is then defined as 
$$\text{CTS} = 2 \cdot \max\left( \frac{D_R}{D_R + D_T} - 0.5, 0 \right),$$
which normalizes the ratio between $D_T$ and $D_R$, with values ranging from 0 to 1. A higher CTS indicates stronger clustering, while if the embeddings are uniformly distributed, $D_R \approx D_T$, making $\frac{D_R}{D_R + D_T} \approx 0.5$, and CTS approaches 0.

\subsection*{Silhouette Index}

The Silhouette Index (SI) is a metric that measures the quality of clustering by evaluating how similar each data point is to its own cluster compared to other clusters. For each data point $i$, the average distance $a(i)$ is computed between $i$ and all other points within the same cluster, which measures the cohesion or how tightly the point is clustered with its own group. Simultaneously, the average distance $b(i)$ is calculated between $i$ and all points in the nearest cluster that $i$ does not belong to, representing the separation or how distinct the cluster is from others.  The Silhouette value for each data point is then calculated using the formula 
$$
s(i) = \frac{b(i) - a(i)}{\max(a(i), b(i))},
$$
where the Silhouette score ranges from $-1$ to $1$. 

A score close to $1$ indicates that the point is well-clustered, a score around $0$ suggests that the point lies near the boundary between two clusters, and a negative score implies that the point might be misclassified, being closer to a different cluster than its own. The overall Silhouette Index is the mean of the silhouette values across all data points, providing an aggregate measure of clustering quality.

\subsection*{Calinski–Harabasz Index}

The Calinski–Harabasz Index (CHI), also known as the Variance Ratio Criterion, is a metric used to assess the quality of clustering by evaluating the ratio of between-cluster dispersion to within-cluster dispersion. 

The between-cluster dispersion $B$ measures how far apart the cluster centroids are from the overall centroid of the data and is calculated as 
$$
B = \sum_{k=1}^K n_k \cdot \|c_k - c\|^2,
$$
where $n_k$ is the number of points in cluster $k$, $c_k$ is the centroid of cluster $k$, and $c$ is the centroid of the entire dataset. 

The within-cluster dispersion $W$ assesses how close the data points are to their respective cluster centroids and is calculated as 
$$
W = \sum_{k=1}^K \sum_{i \in C_k} \|x_i - c_k\|^2,
$$
where $x_i$ is a data point in cluster $k$, and $C_k$ is the set of points in cluster $k$. 

The Calinski–Harabasz Index is then computed as
$$
CHI = \frac{B / (K - 1)}{W / (N - K)},
$$
where $K$ is the number of clusters and $N$ is the total number of data points. 

A higher CHI value indicates that the clusters are well-separated and internally compact, signifying a better clustering structure.

\end{document}